\documentclass[aps,prl,twocolumn,superscriptaddress,amsmath,showpacs]{revtex4}

\usepackage{graphicx, enumitem, multirow}

\begin{document}

\title{Fundamental Structural Constraint of Random Scale-Free Networks}

\author{Yongjoo Baek}
\affiliation {Department of Physics,
Korea Advanced Institute of Science and Technology, Daejeon
305-701, Korea}

\author{Daniel Kim}
\affiliation {Department of Physics,
Korea Advanced Institute of Science and Technology, Daejeon
305-701, Korea}

\author{Meesoon Ha}
\email[Corresponding author; ]{msha@chosun.ac.kr}
\affiliation{Department of Physics Education, Chosun University,
Gwangju 501-759, Korea}

\author{Hawoong Jeong}
\affiliation{Department of Physics and Institute for the
BioCentury, Korea Advanced Institute of Science and Technology,
Daejeon 305-701, Korea} \affiliation{APCTP, Pohang, Gyeongbuk
790-784, Korea}

\date{\today}

\begin{abstract}
We study the structural constraint of random scale-free networks
that determines possible combinations of the degree exponent
$\gamma$ and the upper cutoff $k_c$ in the thermodynamic limit. We
employ the framework of graphicality transitions proposed by [Del
Genio and co-workers, Phys. Rev. Lett. {\bf 107}, 178701 (2011)],
while making it more rigorous and applicable to general values of
$k_c$. Using the graphicality criterion, we show that the upper
cutoff must be lower than $k_c \sim N^{1/\gamma}$ for $\gamma <
2$, whereas any upper cutoff is allowed for $\gamma > 2$. This
result is also numerically verified by both the random and
deterministic sampling of degree sequences.
\end{abstract}

\pacs{89.75.Hc, 02.10.Ox, 89.75.Da, 64.60.aq}


\maketitle
Complex networks~\cite{NetRev} are found in diverse natural and
artificial systems, which consist of heterogeneous elements
(nodes) coupled by connections (links) markedly different from
those of ordinary lattices. In particular, many
systems~\cite{Price1965, Albert1999, Vazquez2002,
HJeong2000,Amaral2000} can be interpreted as scale-free networks,
in which the fraction of nodes with degree $k$ (i.e., $k$ links)
obeys the power-law distribution $P(k) \sim k^{-\gamma}$ over a
broad range of values bounded by $k_m \le k \le k_c$ where
$\gamma$ is called the degree exponent, $k_m$ is the lower cutoff,
and $k_c$ is the upper cutoff. There have been interests in
topological and dynamical properties induced by the degree
distribution, which have been examined through various studies on
random scale-free
networks~\cite{Albert2000,PastorSatorras2001,Sood2008,Dorogovtsev2001,Catanzaro2005,CPq,CPa+Ising_a}.

Random scale-free networks refer to an ensemble of networks
constrained only by the parameters $\gamma$, $k_m$, and $k_c$. In
general, $k_m$ is set as a constant, while $k_c$ is assumed to
increase with the number of nodes $N$ as $k_c \sim N^\alpha$ with
$0 \le \alpha \le 1$. Besides, self-loops or multiple links
between a pair of nodes are often disallowed. Under the
circumstances, the degree exponent $\gamma$ and the cutoff
exponent $\alpha$ determine various properties of networks in the
thermodynamic limit, $N \to \infty$. It is known that $\gamma$
contributes to the resilience against node
failures~\cite{Albert2000}, the epidemic
threshold~\cite{PastorSatorras2001}, the consensus time of opinion
dynamics~\cite{Sood2008}, etc. Meanwhile, $\alpha$ affects the
expected value of the generated maximum
degree~\cite{Dorogovtsev2001}, degree
correlations~\cite{Catanzaro2005}, finite-size scaling at
criticality~\cite{CPq,CPa+Ising_a}, etc.

The studies of random scale-free networks characterized by
$\gamma$ and $\alpha$ must be based on the knowledge that such
networks actually exist in the thermodynamic limit. Hence, it is
necessary to understand the constraint on the possible values of
$\gamma$ and $\alpha$. This problem is exactly equivalent to the
issue of the graphicality of random scale-free networks. A degree
sequence $\left\{k_1, k_2, \ldots, k_N\right\}$ is said to be {\em
graphical} if it can be realized as a network without self-loops
or multiple links. As an indicator of the existence of graphical
sequences, the graphicality fraction $g$~\cite{DelGenio2011} is
defined as the fraction of graphical sequences among the sequences
with an even degree sum generated by $P(k)$. Note that the degree
sequences with an odd degree sum are left out, since such
sequences are trivially nongraphical. The constraint on the
possible values of $\gamma$ and $\alpha$ can be obtained from the
behavior of $g$ due to the fact that the random scale-free
networks with given $\gamma$ and $\alpha$ exist in the
thermodynamic limit if and only if $g$ is nonzero as $N \to
\infty$.
%
Using the graphicality criterion given by the Erd\H{o}s--Gallai
(EG) theorem~\cite{Erdos1960}, Del Genio and
co-workers~\cite{DelGenio2011} have recently studied the behavior
of $g$ as a function of $\gamma$ only for the special case of $k_c
= N - 1$ and $k_m = 1$. They found
\begin{equation*}
g =
\begin{cases}
0 & \text{if $0 \le \gamma \le 2$} \\
1 & \text{otherwise},
\end{cases}
\end{equation*}
where the abrupt changes of $g$ at $\gamma = 0$ and $\gamma = 2$
were termed graphicality transitions~\cite{DelGenio2011}. This
result implies that there exist only sparse random scale-free
networks with a finite average degree ($\gamma > 2$) and
left-skewed networks with an abundance of hubs ($\gamma < 0$) in
the thermodynamic limit when the range of degree is kept maximal.

In this Letter, we generalize their study to arbitrary choices of
degree cutoffs, so that we can provide the complete picture of the
constraint on the possible values of $\gamma$ and $\alpha$.
Starting from the EG theorem, we present a rigorous derivation of
$g$ as a function of both $\gamma$ and $\alpha$, which is then
verified and supplemented by numerical results.
\begin{table}[t]
\centering \caption{\label{tab:k_n} Scalings with $N$ of the $n$th
largest degree $k_n$ for arbitrary values of $\alpha$, $\beta$,
and $\gamma$.}
\begin{tabular}{cccc}
\hline\hline
& $~\beta$ = 0~ & $~0 < \beta < 1~$ & $~\beta = 1~$ \\
\hline
$~\gamma < 1~$ & $~N^\alpha~$ & $~N^\alpha~$ & $~N^\alpha~$ \\
$~\gamma = 1~$ & $~N^\alpha~$ & $~N^\alpha~$ & $~k_m^\nu N^{\alpha(1-\nu)}~$ \\
$~\gamma > 1~$ &
$~N^{\min\left[\alpha,\frac{1}{\gamma-1}\right]}~$ &
$~N^{\min\left[\alpha,\frac{1-\beta}{\gamma-1}\right]}~$ &
$~N^0~$\\
\hline\hline
\end{tabular}
\end{table}
\begin{table}[b]
\centering \caption{\label{tab:scaling} Scalings with $N$ of each
side of the $n$th EG inequality for arbitrary values of $\alpha$,
$\beta$, and $\gamma$.} \scalebox{0.8}{
\begin{tabular}{ccccccccc}
\hline\hline
&&&& $\beta = 0$ && $0 < \beta < 1$ && $\beta = 1$\\
\hline
&& $\gamma < 1$ && $N^\alpha$ && $N^{\alpha+\beta}$ && $N^{\alpha+1}$ \\
&& $\gamma = 1$ && $N^\alpha$ && $N^{\alpha+\beta}$ && $\frac{N^{\alpha+1}}{\ln N}$ \\
lhs && $1 < \gamma < 2$ && $N^\alpha$ && $N^{\min\left[\alpha+\beta,1+\alpha(2-\gamma)\right]}$ && $N^{1+\alpha(2-\gamma)}$ \\
&& $\gamma = 2$ && $N^\alpha$ && $\min\left[N^{\alpha+\beta},N \ln N \right]$ && $N \ln N$ \\
&& $\gamma > 2$ &&
$N^{\min\left[\alpha,\frac{1}{\gamma-1}\right]}$ &&
$N^{\min\left[\alpha+\beta,1-\frac{\gamma-2}{\gamma-1}(1-\beta)\right]}$
&& $N$ \\
\\
&& $\gamma < 1$ && $N$ && $\max\left\{N^{2\beta},N^{1+\min\left[\alpha,\beta\right]}\right\}$ && $N^2$  \\
&& $\gamma = 1$ && $N$ && $\max\left\{N^{2\beta},\min\left[\frac{N^{1+\alpha}}{\ln N}, N^{1+\beta}\right]\right\}$ && $N^2$ \\
rhs && $1 < \gamma < 2$ && $N$ && $\max\left\{N^{2\beta},N^{1+\min\left[\alpha,\beta\right](2-\gamma)}\right\}$ && $N^2$ \\
&& $\gamma = 2$ && $N$ && $\max\left[ N \ln N, N^{2\beta} \right]$ && $N^2$ \\
&& $\gamma > 2$ && $N$ && $\max\left[N, N^{2\beta} \right]$ && $N^2$ \\
\hline\hline
\end{tabular}
}
\end{table}

The EG theorem~\cite{Erdos1960} states that a degree sequence
sorted in the decreasing order $k_1 \ge k_2 \ge \ldots \ge k_N$ is
graphical if it has an even sum and satisfies the EG inequalities
given by
\begin{equation}
\sum_{i=1}^{n} k_i \le n(n-1) + \sum_{i=n+1}^{N} \min \left[n,k_i\right]
\end{equation}
for any integer $n$ in the range $1 \le n \le N-1$.

To determine whether those inequalities are satisfied in the
thermodynamic limit, we derive the network-size scalings of the
left-hand side (lhs) and the right-hand side (rhs) of each
inequality. As the first step, we calculate the scaling of $k_n$
from the cumulative mass function of $k_n$, which is denoted by
$\Gamma^{(n)}_N$. The maximum degree ($n = 1$) satisfies
\begin{equation}
\Gamma^{(1)}_N(k) = \prod_{i=1}^{N}\mathrm{Prob}\left[k_i \le k\right] = \left[C(k)\right]^N
\end{equation}
where $C$ is the cumulative mass function of degree. Since
$\Gamma^{(n)}_N$ satisfies the recursive relation
\begin{align}
\Gamma_N^{(n)}(k) &- \Gamma_N^{(n - 1)}(k) = \mathrm{Prob}\left[\text{$k_{n-1} > k$ and $k_n \le k$}\right] \nonumber \\
&= \binom{N}{n - 1} \left[ 1 - C(k) \right]^{n - 1} \left[ C(k) \right]^{N - n + 1},
\end{align}
we can obtain its exact form as
\begin{equation} \label{eq:cdf1}
\Gamma_N^{(n)}(k) = \sum_{i = 0}^{n - 1} \binom{N}{i} \left[ 1 - C(k) \right]^i \left[ C(k) \right]^{N - i}.
\end{equation}
Suppose $n = \nu N^\beta$, where $\nu > 0$ and $0 \le \beta \le 1$. If $\beta > 0$, we can use the following approximation for large $N$:
\begin{equation} \label{eq:cdf2}
\Gamma^{(n)}_N(k) \approx \frac{\mathrm{erf} \left\{\frac{N\left[C(k)+\nu N^{\beta-1}-1\right]}{\sqrt{2NC(k)\left[1-C(k)\right]}}\right\}+1}{2}.
\end{equation}
Using Eq.~(\ref{eq:cdf1}) for $\beta = 0$ and Eq.~(\ref{eq:cdf2})
for $\beta > 0$, we can find the range of $k$ in which
$\Gamma^{(n)}_{N}(k)$ increases from $0$ to $1$ in the limit $N
\to \infty$. Since the typical values of $k_n$ must fall within
this range of $k$, we can obtain the network-size scalings of
$k_n$ as listed in Table~\ref{tab:k_n}.

While both sides of the EG inequalities are sums over $k_n$, we
can approximate those sums as integrals, since it does not affect
the the leading $N$-dependent term that determines the scaling
relation. It is straightforward to approximate the lhs, while the
rhs needs a careful reformulation. The second term of the rhs
satisfies
\begin{align}
\sum_{i=n+1}^{N} \min &\left[n,k_i\right] = N\theta\left(n-k_m\right)\sum_{k=k_m}^{\min\left[n,k_{n+1}\right]}kP(k) \nonumber \\
&+ N\theta\left(k_{n+1}-n-1\right)\sum_{\max\left[n+1,k_m\right]}^{k_{n+1}}nP(k)
\end{align}
where $\theta$ denotes the Heaviside step function defined by
$\theta(x) = 1$ if $x \ge 0$, and $\theta(x) = 0$ otherwise. From
now on, each summation can be converted to an integral over the
same range. Calculating all the integrals, we can single out the
leading $N$-dependent terms of each side, as listed in
Table~\ref{tab:scaling}. The scalings of those terms are
completely determined by the three exponents $\alpha$, $\beta$,
and $\gamma$, while the lower cutoff $k_m$ turns out to be
irrelevant.

We can now determine whether the EG inequalities are satisfied
through the comparison of scaling exponents in both sides. By the
EG theorem, $g = 1$ if the inequalities are satisfied for all
possible values of $\beta$, and $g = 0$ if there exist the values
of $\beta$ at which some inequalities are violated. Hence, the
asymptotic behavior of $g$ is obtained as follows:
\begin{equation} \label{eq:g}
g =
\begin{cases}
0 & \text{if $1/\alpha < \gamma < 2$} \\
1 & \text{if $\gamma > 2$ or $\alpha < \min\left[1/\gamma,1\right]$}.
\end{cases}
\end{equation}

We note that the behavior of $g$ for the special case of $\alpha =
1$ and $\gamma < 1$ cannot be determined by our scaling argument,
since both sides of the EG inequalities satisfy the same
network-size scalings. To address this problem analytically, it is
necessary that we consider the coefficients of the leading
$N$-dependent terms, which is beyond the scope of this Letter.
Instead, we settle for its numerical resolution at the end of
this Letter.

For the other cases, we can analytically determine the locations
of graphicality transitions from Eq.~(\ref{eq:g}). There exist two
transition points for each value of $\alpha$ in the range $1/2 <
\alpha < 1$, namely the upper transition point
$\gamma^*_\mathrm{U} = 2$ and the lower transition point
$\gamma^*_\mathrm{L} = 1/\alpha$. On the other hand, no transition
occurs for $0 \le \alpha \le 1/2$ where $g = 1$ always holds.
\begin{figure}[b]
\centering
\includegraphics[width=0.95\columnwidth]{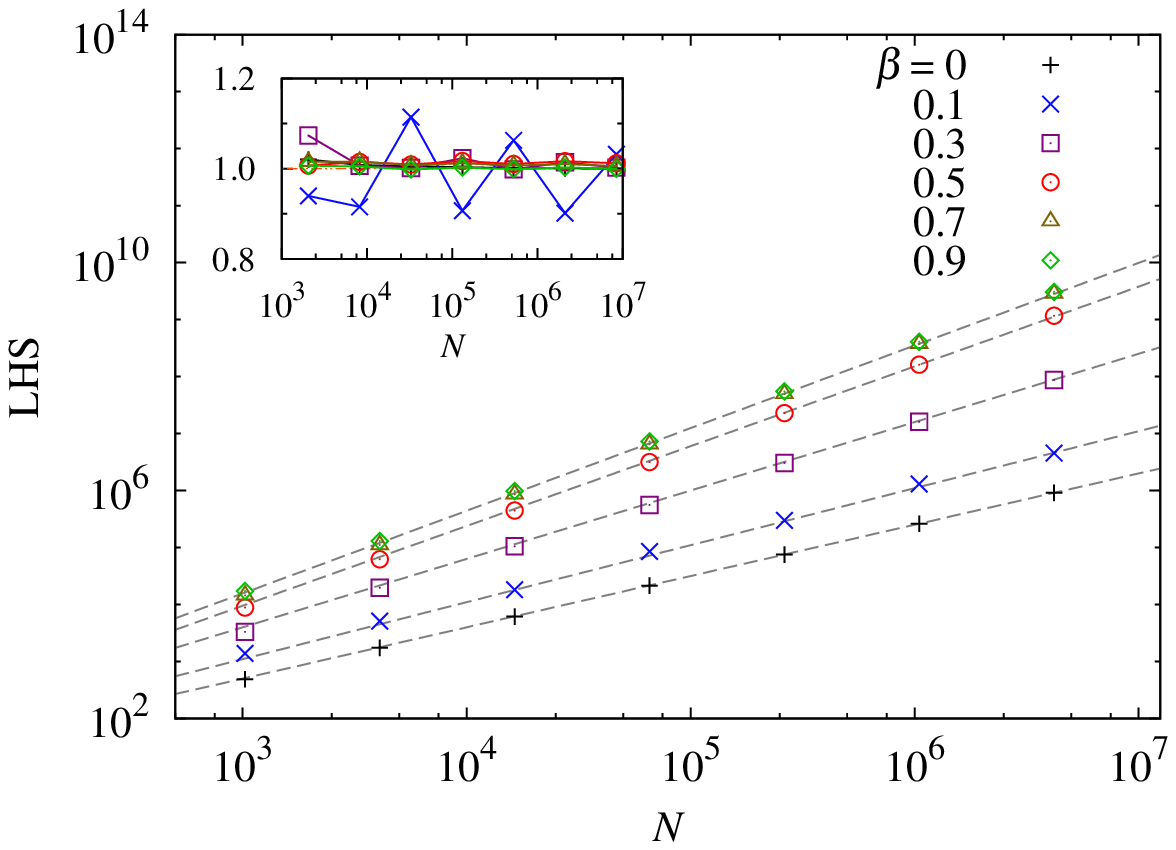} \\
\includegraphics[width=0.95\columnwidth]{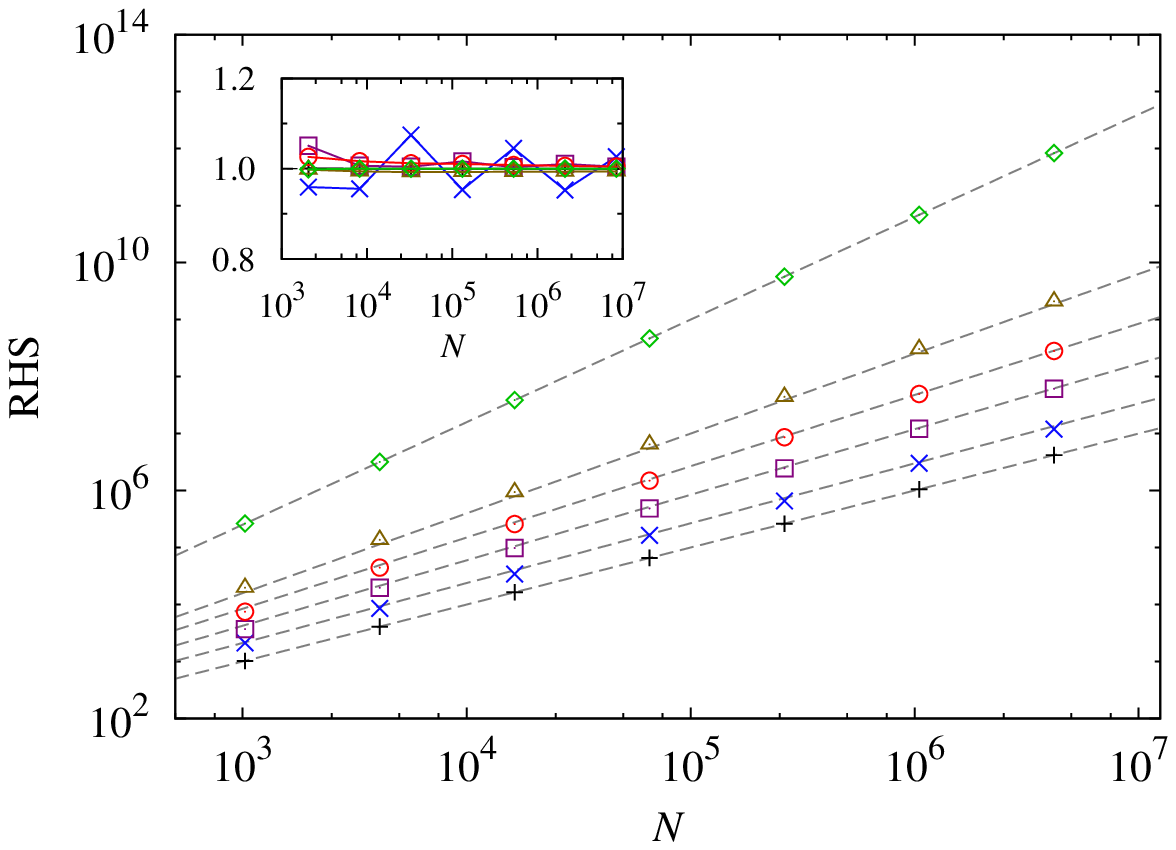}
\caption{\label{fig:scaling} (Color online) $N$ dependence of the
$n$th EG inequality obtained from $10^3$ random degree sequences
(symbols) for $\gamma=1.5$ and
 $\alpha=0.9$. It is in good agreement with the scalings (lines)
predicted in Table~\ref{tab:scaling}, which is also confirmed by
the insets: The ratio between the successive slopes of each symbol
and the slope of the corresponding line stays close to one.}
\end{figure}
\begin{figure}[t]
\centering
\includegraphics[width=0.975\columnwidth]{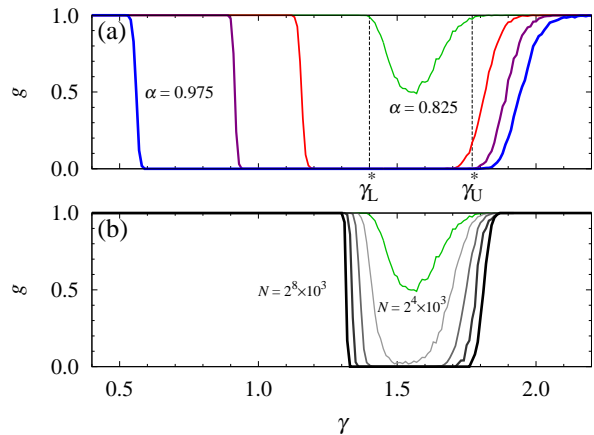}
\caption{\label{fig:g_gamma} (Color online) $\gamma$ dependence of
the graphicality fraction $g$ obtained from $10^3$ random degree
sequences. (a) $\alpha$ increases by steps of $0.05$ with $N = 2^4
\times 10^3$. $\gamma^*_\mathrm{L}$ and $\gamma^*_\mathrm{U}$
indicate the two transition points at $\alpha = 0.825$, whose $N$
dependence is shown in (b) as $N$ increases by factors of $2$. }
\end{figure}

All the predictions on the asymptotic behavior of $g$ can be
numerically checked by the evaluation of the EG inequalities.
First of all, we can verify all the scalings listed in
Table~\ref{tab:scaling}, some of which are shown in
Fig.~\ref{fig:scaling}. This indirectly supports our predictions
on the behavior of $g$, as all the predictions were deduced from
those scalings.

To obtain direct support for our predictions, we need to measure
the $\gamma$ dependence of $g$ from the random samples of degree
sequences, as illustrated in Fig.~\ref{fig:g_gamma}. Due to
sample-to-sample fluctuations at finite system size, $g$ changes
continuously between $0$ and $1$ over a finite range of $\gamma$,
which becomes narrower as $N$ increases. In fact, $g$ may not even
reach $0$ if $N$ is too small, as exemplified by the curves for
$\alpha = 0.825$ in Fig.~\ref{fig:g_gamma}(a). The curves in
Fig.~\ref{fig:g_gamma}(b) show that the minimum of $g$ gradually
reaches zero as $N$ increases. Keeping those observations in mind,
for the sake of convenience, we regard the range of $\gamma$ in
which $g$ falls below $0.99$ as the effectively nongraphical
region. Then, the boundary of this region can be chosen as the
effective transition points at finite $N$, which are again marked
as $\gamma^*_\mathrm{L}$ and $\gamma^*_\mathrm{U}$ in
Fig.~\ref{fig:g_gamma}(a).

\begin{figure}[b]
\centering
\includegraphics[width=0.9\columnwidth]{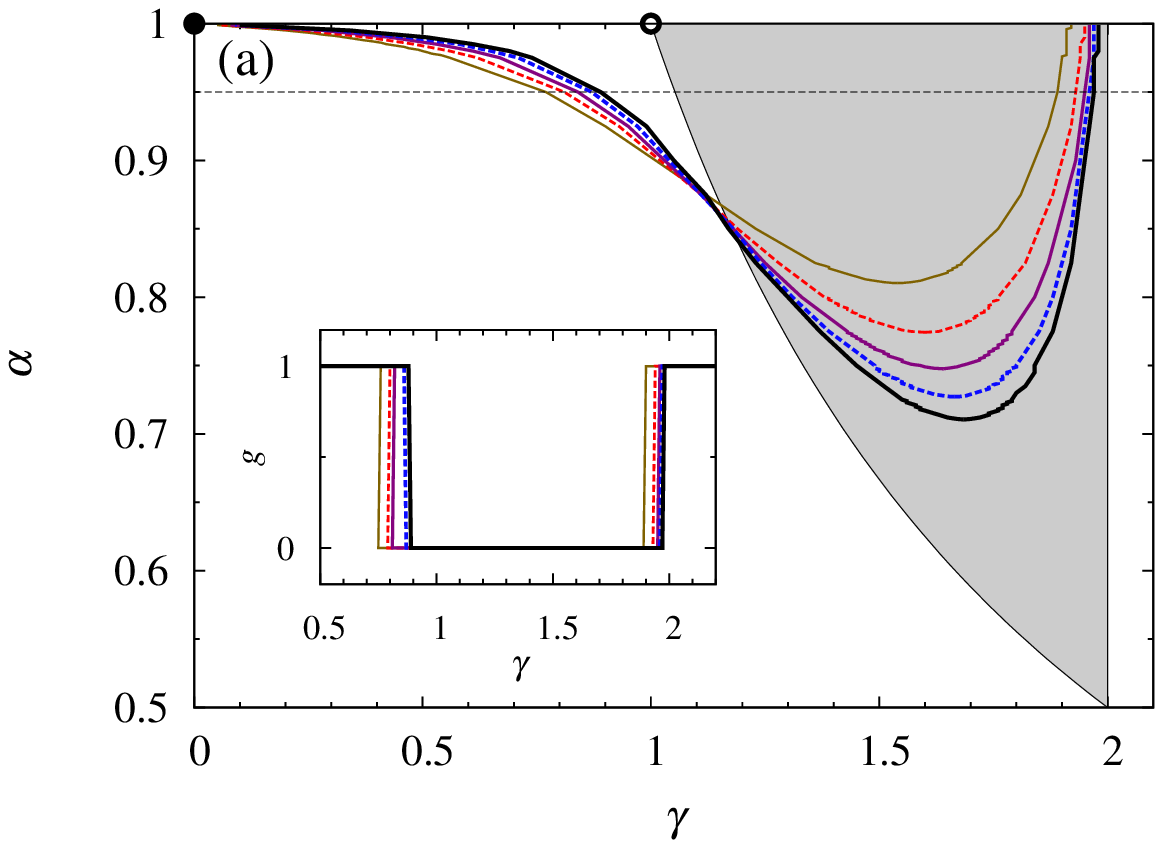} \\
\includegraphics[width=0.9\columnwidth]{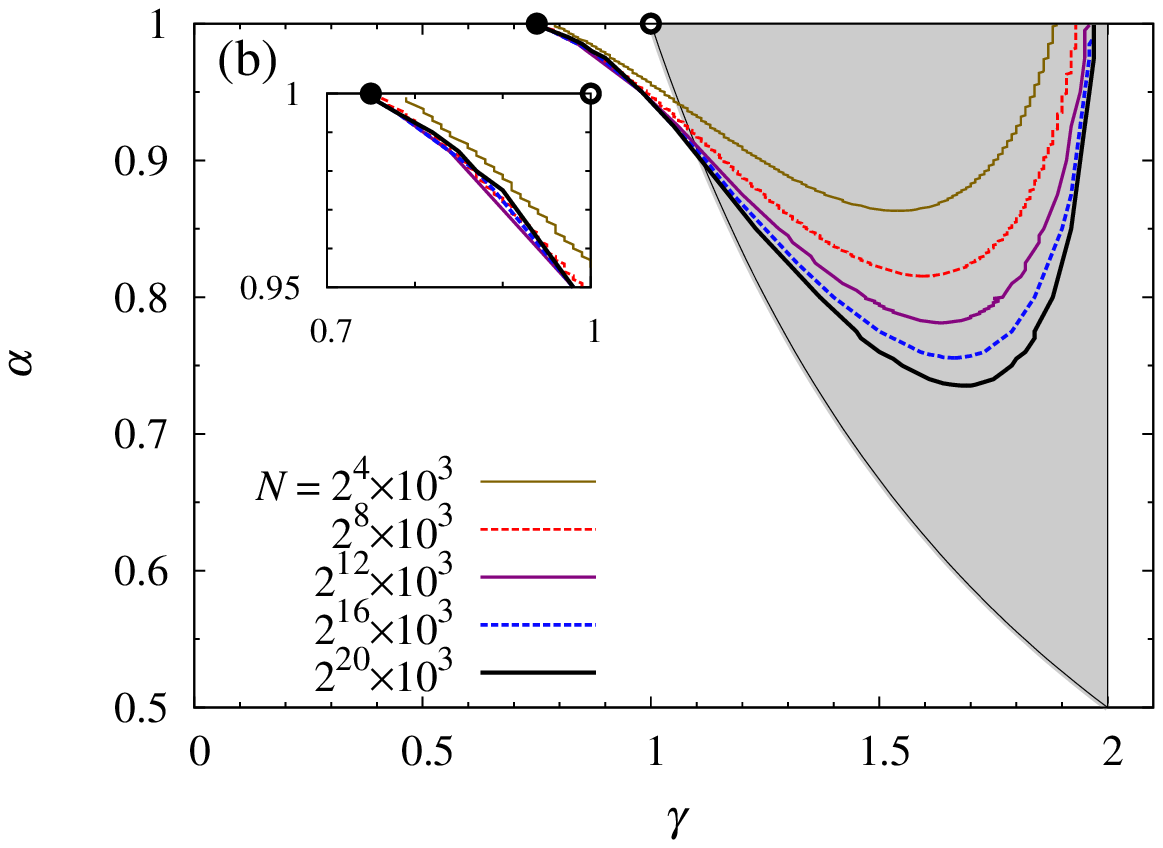}
\caption{\label{fig:gamma_alpha} (Color online) Graphicality
diagram in the $(\gamma,\alpha)$ plane at (a) $c = 1$ and (b) $c =
0.6$. The nongraphical region is in gray, while the other lines
are the transition lines estimated using deterministic degree
sequences, whose behavior of $g$ is as illustrated for $\alpha =
0.95$ in the inset of (a).}
\end{figure}

\begin{figure}[t]
\centering
\includegraphics[width=0.9\columnwidth]{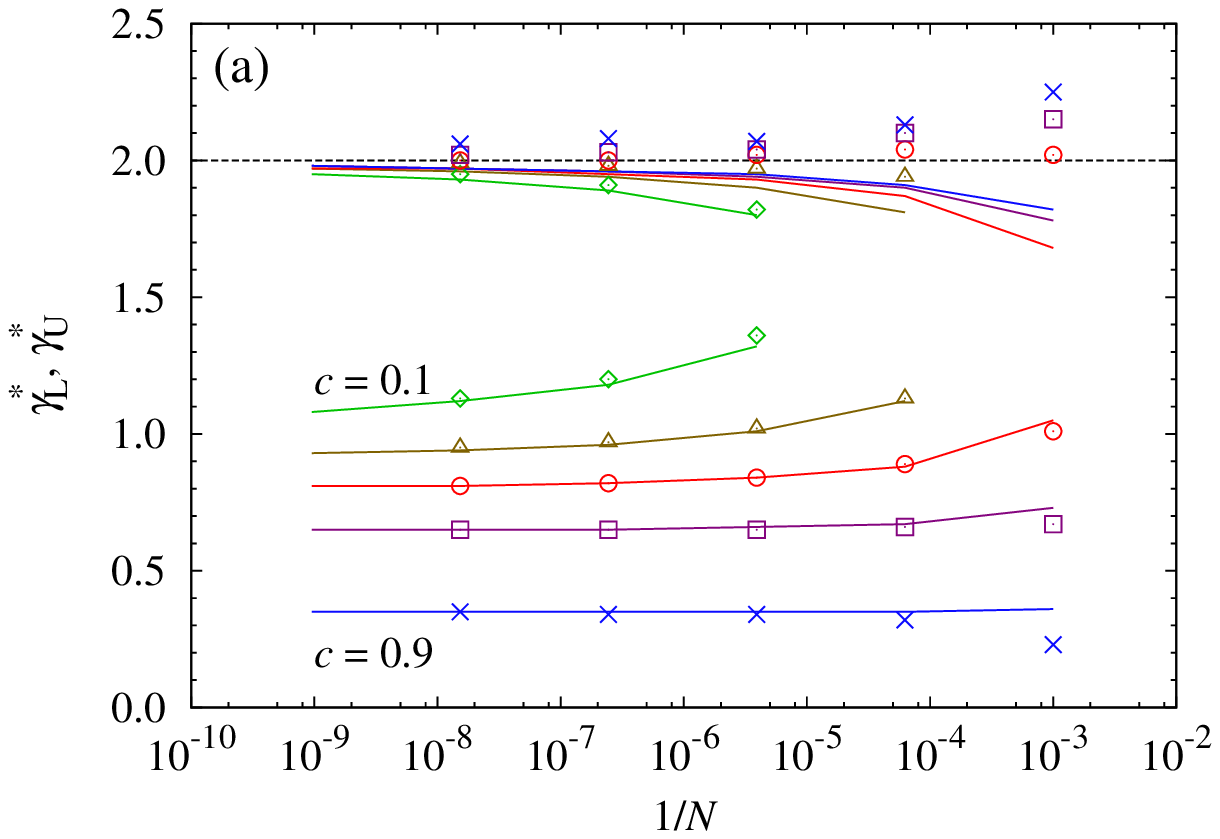} \\
\includegraphics[width=0.9\columnwidth]{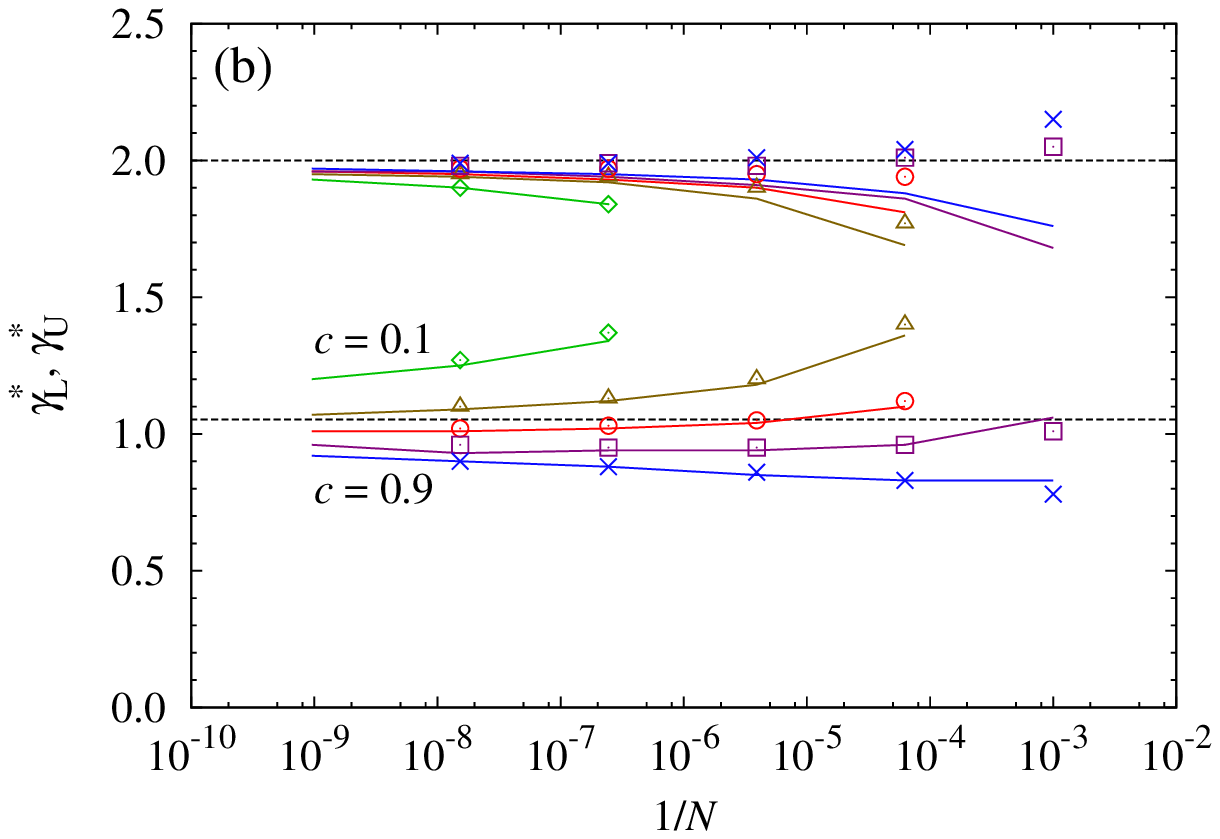}
\caption{\label{fig:gamma*_N} (Color online) $N$ dependence of the
transition points as $c$ increases by steps of 0.2 for (a) $k_c =
c(N-1)$ and (b) $k_c = cN^{0.95}$. Symbols indicate the transition
points obtained from up to $10^3$ random degree sequences, while
those obtained from deterministic sequences are connected with
lines.}
\end{figure}

We also consider deterministically generated degree sequences
defined by $1 - C(k_n) = \nu$, obtained from Eq.~(\ref{eq:cdf2}),
which ensures that the sequences exactly follow the network-size
scalings of $k_n$ for $n = \nu N$. Those sequences filter out the
sample-to-sample fluctuations, making it very straightforward to
locate the effective transition points [see the inset
Fig.~\ref{fig:gamma_alpha}(a)]. They also greatly improve the
efficiency of calculation, allowing us to check our predictions at
larger $N$. We observe that the transition points estimated by
both randomly and deterministically generated degree sequences
approach each other as $N \to \infty$ (for example, see
Fig.~\ref{fig:gamma*_N}). Therefore, we can use either of those
two different samplings to numerically check our predictions.

In Fig.~\ref{fig:gamma_alpha}, we present graphicality diagrams
obtained at two different values of the cutoff coefficient $c$,
where the transition lines at finite network sizes are estimated
by the deterministic samplings of degree sequences, and also
compared with the transition lines in the thermodynamic limit
predicted by the scaling argument. The numerically estimated
transition lines tend to approach analytically predicted ones,
which are independent of $c$. Combining this observation with the
verification of the scaling relations listed in
Table~\ref{tab:scaling}, we can safely conclude that the numerical
results are slowly converging to our predictions as $N$ increases.

Moreover, Fig.~\ref{fig:gamma_alpha} gives us some clues as to the
location of the graphicality transitions for $\alpha = 1$ and
$\gamma < 1$, which could not be determined by the scaling
argument as previously explained. The estimated transition lines
suggest that the location of the transition is dependent on $c$:
$\gamma^*_\mathrm{L}$ approaches $\gamma = 0$ at $c = 1$, as
previously reported~\cite{DelGenio2011}, but it converges to some
different limiting value if $c \ne 1$. The effect of $c$ on the
graphicality at $\alpha = 1$ is more closely examined in
Fig.~\ref{fig:gamma*_N}(a), which suggests that
$\gamma^*_\mathrm{L}$ varies continuously between 0 and 1 with
$c$, while $\gamma^*_\mathrm{U}$ converges to $2$, regardless of
$c$. This nicely contrasts with Fig.~\ref{fig:gamma*_N}(b), which
confirms our prediction that the transition points are independent
of $c$ for $\alpha < 1$.

While we have given an almost complete picture of the graphicality
issue of scale-free networks, the nature of graphicality
transitions requires further studies. Note that at a transition
point, the comparison of scalings fails to determine whether the
Erd\H{o}s--Gallai inequality holds in the asymptotic limit, just
like the case of $\alpha = 1$. In such cases, the coefficients of
the leading-order terms as well as the second- and higher-order
terms must be considered to determine the value of $g$. Thus, we
cannot claim yet that graphicality transitions are truly
discontinuous as previously claimed~\cite{DelGenio2011}, since
they might be sharp but continuous transitions resembling the
continuous change of $\gamma^*_\mathrm{L}$ with $c$ at $\alpha =
1$. The claim should be either proven or disproven by a more
complete understanding of the behavior of $g$ at the transition
points.

In conclusion, we have found that in the thermodynamic limit
random scale-free networks without self-loops or multiple links
are either sparse ($\gamma > 2$) with arbitrary values of degree
cutoffs, or dense ($0 < \gamma < 2$) with the upper cutoff $k_c
\sim N^\alpha$ satisfying $\alpha < 1/\gamma$, supplementing the
statement that ``all (random) scale free networks (with maximal
range of degree) are sparse.''~\cite{DelGenio2011} This also
agrees with the upper cutoff found by Seyed-allaei {\em et
al.}~\cite{Seyedallaei2006}, which is required for scale-free
networks with $\gamma < 2$ generated using a node-fitness
mechanism~\cite{Caldarelli2002}. We also numerically found that
the cutoff coefficient $c$ affects the realizability of degree
sequences for the special case of the linear cutoff $\alpha = 1$,
which has been overlooked. Our results impose a limit on the
values of $\gamma$ and $\alpha$ for which the properties of random
scale-free networks numerically obtained in finite systems can be
extrapolated to the thermodynamic limit.

The work was supported by the National Research Foundation of
Korea (NRF) Grant funded by the Korean Government (MEST) (No.
2011-0011550) (M.H.); (No. 2011-0028908) (Y.B., D.K., H.J.). M.H.
also acknowledges the generous hospitality of KIAS for the
Associate Member Program, funded by the MEST.


\begin{thebibliography}{99}
\bibitem{NetRev} R. Albert and A.-L. Barab\'{a}si, Rev. Mod. Phys. {\bf 74}, 47 (2002);
S.N. Dorogovtsev and J.F.F. Mendes, Adv. Phys. {\bf 51}, 1079
(2002); M.E.J. Newman, SIAM Rev. {\bf 45}, 167 (2003); S.
Boccaletti, V. Latora, Y. Moreno, M. Chavez, and D.-U. Hwang,
Phys. Rep. {\bf 424}, 175 (2006); S.N. Dorogovtsev, A.V. Goltsev,
and J.F.F. Mendes, Rev. Mod. Phys. {\bf 80}, 1275 (2008).

\bibitem{Price1965} D.J. de Solla Price, Science {\bf 149}, 510 (1965).
\bibitem{Albert1999} R. Albert, H. Jeong, and A.-L. Barab\'{a}si, Nature (London) {\bf 401}, 130 (1999).
\bibitem{Vazquez2002} A. V\'{a}zquez, R. Pastor-Satorras, and A. Vespignani, Phys. Rev. E {\bf 65}, 066130 (2002).
\bibitem{HJeong2000} H. Jeong, B. Tombor, R. Albert, Z. N. Oltvai, and A.-L. Barab\'{a}si, Nature (London) {\bf 407}, 651 (2000).
\bibitem{Amaral2000} L.A.N. Amaral, A. Scala, M. Barthelemy,  and H.E. Stanley, Proc. Natl. Acad. Sci. U.S.A. {\bf 97}, 11149 (2000).

\bibitem{Albert2000} R. Albert, H. Jeong, and A.-L. Barab\'{a}si, Nature (London) {\bf 406}, 378 (2000).
\bibitem{PastorSatorras2001} R. Pastor-Satorras and A. Vespignani, Phys. Rev. E {\bf 63}, 066117 (2001).
\bibitem{Sood2008} V. Sood, T. Antal, and S. Redner, Phys. Rev. E {\bf 77}, 041121 (2008).

\bibitem{Dorogovtsev2001} S.N. Dorogovtsev, J.F.F. Mendes, and A.N. Samukhin, Phys. Rev. E {\bf 63}, 062101 (2001).
\bibitem{Catanzaro2005} M. Catanzaro, M. Bogu\~{n}\'{a}, and R. Pastor-Satorras, Phys. Rev. E {\bf 71}, 027103 (2005).
\bibitem{CPq} C. Castellano and R. Pastor-Satorras, Phys. Rev. Lett. {\bf 96}, 038701
(2006); M. Ha, H. Hong, and H. Park, {\em ibid.} {\bf 98}, 029801
(2007); C. Castellano and R. Pastor-Satorras, {\em ibid.} {\bf
98}, 029802 (2007); H. Hong, M. Ha, and H. Park, {\em ibid.} {\bf
98}, 258701 (2007).
\bibitem{CPa+Ising_a} C. Castellano and R. Pastor-Satorras, Phys. Rev. Lett. {\bf 100},
148701 (2008); M. Bogu\~{n}\'{a}, C. Castellano, and R.
Pastor-Satorras, Phys. Rev. E {\bf 79}, 036110 (2009); J.D. Noh
and H. Park, {\em ibid.}, {\bf 79}, 056115 (2009); S.H. Lee, M.
Ha, H. Jeong, J.D. Noh, and H. Park, {\em ibid.} {\bf 80}, 051127
(2009).

\bibitem{DelGenio2011} C.I. Del Genio, T. Gross, and K.E. Bassler, Phys. Rev. Lett. {\bf 107}, 178701 (2011).
\bibitem{Erdos1960} P. Erd\H{o}s and T. Gallai, Matematikai lapok {\bf 11}, 264 (1960).

\bibitem{Seyedallaei2006} H. Seyed-allaei, G. Bianconi, and M. Marsili, Phys. Rev. E {\bf 73}, 046113 (2006).
\bibitem{Caldarelli2002} G. Caldarelli, A. Capocci, P. De Los Rios, and M.A. Mu\~{n}oz,
Phys. Rev. Lett. {\bf 89}, 258702 (2002).



\end{thebibliography}
\end{document}